\documentclass[twocolumn,prl,superscriptaddress,noshowpacs,epsf]{revtex4}

\usepackage{graphicx}

\begin{document}
\title{Simultaneous cooling of an artificial atom and its neighboring 
quantum system}
\date{\today}
\author{J. Q. You}
\affiliation{Department of Physics and Surface Physics Laboratory (National Key
Laboratory), Fudan University, Shanghai 200433, China}
\affiliation{Frontier Research System, The Institute of Physical
and Chemical Research (RIKEN), Wako-shi 351-0198, Japan}
\author{Yu-xi Liu}
\affiliation{Frontier Research System, The Institute of Physical
and Chemical Research (RIKEN), Wako-shi 351-0198, Japan}
\affiliation{CREST, Japan Science and Technology Agency (JST),
Kawaguchi, Saitama 332-0012, Japan}

%
\author{Franco Nori}
\affiliation{Frontier Research System, The Institute of Physical
and Chemical Research (RIKEN), Wako-shi 351-0198, Japan}
\affiliation{CREST, Japan Science and Technology Agency (JST),
Kawaguchi, Saitama 332-0012, Japan}
\affiliation{Center for Theoretical Physics, Physics Department,
Center for the Study of Complex Systems,
University of Michigan, Ann Arbor, MI 48109-1040, USA}

\begin{abstract}
We propose an approach for cooling both an artificial atom (e.g., a flux qubit) 
and its neighboring quantum system, the latter modeled by either a quantum 
two-level system or a quantum resonator. The flux qubit is cooled by manipulating 
its states, following an inverse process of state population inversion, 
and then the qubit is switched on to resonantly interact with the 
neighboring quantum system. By repeating these steps,  
the two subsystems can be simultaneously cooled. Our results show that 
this cooling is robust and effective, irrespective of the chosen quantum systems 
connected to the qubit.

\end{abstract}
\maketitle

Quantum devices using Josephson junctions can be used as artificial atoms (AAs) 
for demonstrating quantum phenomena at macroscopic scales. With states involving 
the two lowest energy levels, these devices are good candidates for solid-state 
qubits~\cite{YN05}. When using their three lowest levels, 
such a solid-state three-level system, fabricated on a microelectronic chip, 
can be useful for single-photon production~\cite{micromaser} and lasing~\cite{laser}.

\begin{figure}
\includegraphics[width=3.4in,  
bbllx=120,bblly=266,bburx=570,bbury=602]{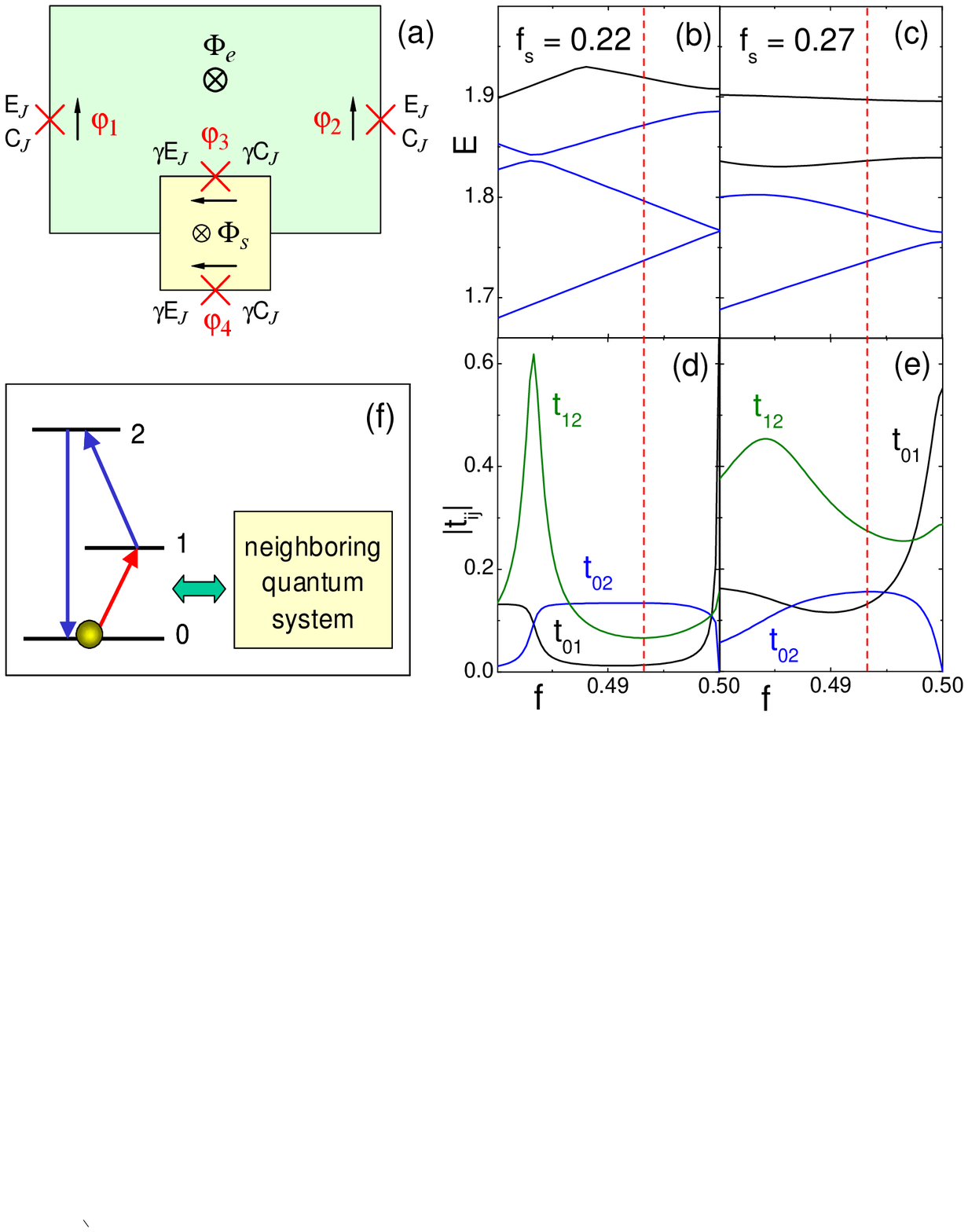} \caption{(Color
online) (a)~Schematic diagram of an artificial atom (AA) produced by
a superconducting quantum circuit. A symmetric SQUID and two identical Josephson 
junctions with coupling energy $E_J$ and capacitance $C_J$ are placed in a
superconducting loop pierced by a magnetic flux $\Phi_e$ (green). 
The two junctions in the SQUID have coupling energy $\gamma E_J$ and capacitance
$\gamma C_J$, and the flux (yellow) threading through the SQUID loop is $\Phi_s$. 
Here $\gamma=0.5$ and $E_c=e^2/2C_J$ is the single-particle
charging energy of the junction. 
(b) and (c):~Energy levels of the superconducting AA as a function of the 
reduced magnetic flux 
$f=\Phi_e/\Phi_0+f_s/2$, for $f_s\equiv \Phi_s/\Phi_0=0.22$ and $0.27$, 
where only the four lowest levels are shown and the energy is in units of $E_J$. 
(d) and (e):~Moduli of the transition matrix elements $|t_{ij}|$ 
(in units of $I_c\Phi_w^{(0)}$) as a function of $f$, for $f_s=0.22$ and $0.27$. 
Note that each figure in (b)-(e) is symmetric about $f=0.5$ and half of it is plotted.
The vertical (red) dashed lines at $f=0.493$ are just a guide to the eye. 
(f)~Transition diagram of the AA. 
At nonzero temperatures, the flux qubit is thermally
activated from the ground state $|0\rangle$ to the first excited state $|1\rangle$.
A resonant transition from $|1\rangle$ to the second excited state 
$|2\rangle$ is driven by a microwave field, so as to eliminate the unwanted thermal 
population of $|1\rangle$,
and followed by a fast decay to $|0\rangle$. While the qubit is cooled to 
its ground state $|0\rangle$, the AA is then switched on, to resonantly interact with 
a neighboring quantum system 
for a period of time. 
Repeating these processes, both the qubit and the neighboring quantum system 
can be simultaneously cooled.   
} \label{fig1}
\end{figure}

For single-photon production~\cite{micromaser} and AA lasing~\cite{laser}, 
a state population inversion is established for the two working energy levels 
via a third one (i.e., transitions $|0\rangle\rightarrow |2\rangle\rightarrow 
|1\rangle$ in Fig.~\ref{fig1}). Interestingly, the inverse process of state 
population inversion (i.e., transitions $|1\rangle\rightarrow |2\rangle\rightarrow 
|0\rangle$ in Fig.~\ref{fig1}) can be used to increase the 
occupation probability of the ground state and thus lower the temperature 
of the qubit. This idea has been applied in a recent experiment~\cite{MIT} 
to cool a flux qubit. Indeed, this is analogous to the optical side-band cooling 
method studied earlier (see, e.g., \cite{sideband} and \cite{sideband1}). 
The experiment~\cite{MIT} shows that the temperature of the flux qubit can be 
lowered by up to two orders of magnitude with respect to its surroundings. 
This provides an efficient approach for preparing a flux qubit in its ground state. 

While the flux qubit was greatly cooled in \cite{MIT}, the noise sources 
surrounding the qubit were not. This is because of 
the weak coupling between the qubit and its environment in \cite{MIT}, 
where the transition rate between the ground and first excited states is small. 
Below we use a tunable AA (to be specific, we choose a flux qubit, but it could be 
another AA) to achieve a strong and switchable 
coupling between the AA and its neighboring quantum system, and propose an approach 
to simultaneously cool both of them and not just the AA.       
Here we consider two typical quantum systems to describe the environment 
surrounding the AA:~(i)~a quantum two-level system (TLS), which is exactly solvable; 
and (ii)~a quantum resonator. Actually,
a quantum TLS can describe the noise source like a two-level fluctuator and 
the quantum resonator can model the dominant bosons of a thermal bath. In this case, 
the approach is to cool both the flux qubit and such noise sources. 
This simultaneous cooling of the flux qubit and its neighboring noise sources 
can significantly enhance the quantum coherence of the flux qubit because 
the cooled qubit is thermally activated very slowly to the first excited state, 
after its neighboring noise sources are also cooled. 
Moreover, the present approach has wide applications because the models  
used here can describe other quantum systems. Also, we show that different 
surrounding quantum systems (either a quantum TLS or a quantum resonator)
give similar results, implying that the cooling is  
robust and effective, irrespective of the chosen neighboring quantum system. 

{\it Cooling the artificial atom and ground-state preparation.}{\bf---}The commonly 
used flux qubit~\cite{FLUX,LIU} (which is an example for an AA) consists of a 
superconducting loop interrupted by three Josephson junctions 
(two equal and one smaller) and pierced by a magnetic flux $\Phi_e$. 
To obtain a tunable AA, the smaller junction is here 
replaced by a SQUID threaded by a flux $\Phi_s$ [see Fig.~\ref{fig1}(a)]. 
The Hamiltonian can be written as
$H={P_p^2}/{2M_p}+{P_q^2}/{2M_q}+U(\varphi_p,\varphi_q)$,
with $P_i=-i\hbar\partial/\partial\varphi_i$ ($i=p,q$),
$M_p=2C_J(\Phi_0/2\pi)^2$, and $M_q=M_p(1+4\gamma)/4$.
The potential 
is 
$U(\varphi_p,\varphi_q)=2E_J[1-\cos\varphi_p\cos(\pi f+\frac{1}{2}\varphi_q)]
+2\gamma E_J[1-\cos(\pi f_s)\cos\varphi_q)$, 
where $\varphi_p=(\varphi_1+\varphi_2)/2$, $\varphi_q=(\varphi_3+\varphi_4)/2$, 
$f_s=\Phi_s/\Phi_0$, and $f=\Phi_e/\Phi_0 + f_s/2$ ($\Phi_0$ is the flux quantum).
To drive a resonant transition between states $|E_i\rangle$ and $|E_j\rangle$, one 
can apply a microwave field through the circuit loop:
$\Phi_w(t)=\Phi_w^{(0)}\cos(\omega_{ij}t+\theta)$, with $\omega_{ij}=(E_i-E_j)/\hbar$. 
When the microwave field is weak, 
the time-dependent perturbation Hamiltonian can be written as
$V(t)=-I\Phi_w(t)$, where
$I=-I_c\cos\varphi_p\sin(\pi f+\frac{1}{2}\varphi_q)$, with $I_c=2\pi E_J/\Phi_0$. 
The rate of the state transition between $|E_i\rangle$ and $|E_j\rangle$ is 
$\Gamma_{ij}\propto |t_{ij}|^2$, where $t_{ij}=\langle E_i|I\Phi_w^{(0)}|E_j \rangle$ 
is the transition matrix element. When a neighboring quantum system, e.g., a noise 
source, is coupled to the flux qubit via a flux variation, then $\Phi_w^{(0)}$ 
in $t_{ij}$ becomes the amplitude of the flux variation.

In Figs.~\ref{fig1}(b)-\ref{fig1}(e), we show the energy levels of the AA and the 
transition matrix elements $|t_{ij}|$ for two values of $f_s$. The SQUID gives 
an effective Josephson coupling energy $\alpha E_J$ with $\alpha=0.77$ and $0.66$,
respectively. For any nonzero temperature, 
the system will be thermally activated 
from the ground state $|0\rangle\equiv |E_0\rangle$ to the first excited state 
$|1\rangle\equiv |E_1\rangle$. 
Here we consider the case in Fig.~\ref{fig1}(b), with the system working at, e.g., 
$f=0.493$. As shown in Fig.~1(d), at this $f$, the corresponding transition matrix 
elements are 
$|t_{01}|\approx 0.01$, $|t_{12}|\approx 0.07$, and $|t_{02}|\approx 0.13$. When a 
microwave field is applied to drive a resonant transition 
$|1\rangle\longrightarrow |2\rangle\equiv |E_2\rangle$, 
because $\Gamma_{20}>\Gamma_{21}\gg \Gamma_{10}$ at $f\sim 0.493$, 
the system can be pumped 
from $|1\rangle$ to $|2\rangle$ and then quickly decays to the ground state 
$|0\rangle$, while the process for thermally activating the system from 
$|0\rangle$ to $|1\rangle$, via coupling to the environment, will be very slow.
Note that the coupling strength of the states $|2\rangle$ and $|0\rangle$ to 
the flux noise source is also proportional to the transition matrix element 
$|t_{20}|$, so the decay rate from $|2\rangle$ to $|0\rangle$ is proportional to 
$\Gamma_{20}$~($\propto |t_{20}|^2$) according to the Fermi golden 
rule~\cite{Cohen}. Therefore, 
the flux qubit is 
``cooled'' because the population probability for the ground state $|0\rangle$ 
can be greatly increased, with respect to any unwanted excited state $|1\rangle$. 
Interestingly, this cooling mechanism corresponds to an ``inverse process" 
of the usual state population inversion. 
For simplicity, here we use a weak microwave field. The driving field 
would need to be stronger to achieve cooling when the relevant transition matrix 
elements are small. This puts some constraints on the specific 
amplitudes used to achieve the desired result~\cite{MIT-NEC}. 
Indeed, a recent experiment~\cite{MIT} has successfully 
realized the microwave-induced cooling, lowering the temperature of a flux qubit 
relative to its surroundings. Thus, this microwave-induced cooling provides 
an efficient method for preparing the flux qubit in its ground state.  
Below we use this prepared ground state to further cool a quantum system 
connected to the qubit.

{\it Cooling a quantum two-level system.}{\bf---}In the subspace spanned by 
$|0\rangle$ and $|1\rangle$, the flux qubit (our AA) is modeled by 
$H_q=\frac{1}{2}\hbar\omega_{10}\sigma_z$. Here we consider a qubit-TLS system 
described by $H_t=H_q+H_{\rm TLS}+V+H_{\rm env}$, 
where $H_{\rm TLS}=\frac{1}{2}\hbar\Omega\sigma'_z$
is the Hamiltonian of a quantum TLS and $H_{\rm env}$ describes all the 
degrees of freedom in the environment and their coupling to the TLS. 
Hereafter, the Pauli operators with primes refer to the neighboring TLS.
The interaction Hamiltonian between the qubit and the TLS
is $V=\hbar g(\sigma_{+}\sigma'_{-}+{\rm H.c.})$, with $g=|t_{01}|/\hbar$. 
In the experimental case~\cite{MIT}, corresponding to Fig.~\ref{fig1}(d), 
because $|t_{01}|$ 
is small at $f\sim 0.493$, the coupling between the qubit and its environment is weak. 
To cool the TLS effectively, after the qubit with $f_s=0.22$ is cooled to the ground 
state, 
we change the reduced magnetic flux $f_s$ to $f_s=0.27$, which corresponds to 
Fig.~\ref{fig1}(e). 
For the qubit parameters used here, 
it is shown~\cite{micromaser} that at $f\sim 0.493$ (i.e., in between 
the level-crossing points), the adiabatic condition
$|{\hbar\langle E_i|(d/dt)|E_j\rangle}/(E_i-E_j)|\ll 1$
can still be fulfilled for the three lowest levels 
by changing the applied flux as fast as $0.1\,\Phi_0$~ns$^{-1}$. 
This means that around this $f$ the quantum states can be well preserved 
even when changing the flux very fast. More importantly, in the case 
of Fig.~\ref{fig1}(e), because $|t_{01}|$ is much increased, then the qubit-TLS 
interaction $\hbar g$ is strengthened by one order of magnitude. 
Here we assume that the quantum TLS is resonant to the qubit with $f_s=0.27$. 
Since the level spacing $\hbar\omega_{01}$ of the qubit with $f_s=0.22$ is different 
from that with $f_s=0.27$, thus at $f\sim 0.493$ the qubit with $f_s=0.22$ 
is off-resonant to the TLS. This gives an even smaller effective qubit-TLS coupling. 

For simplicity, we now assume that the flux qubit is ideally cooled to the 
ground state $|0\rangle$ and then begins to resonantly interact 
with the quantum TLS at time $t_i$. When $H_{\rm env}$ is not included, 
the time evolution of the density operator of the TLS is governed by 
$\rho(t_i+\tau)=M(\tau)\rho(t_i)$ and the gain operator is defined by 
$M(\tau)\rho={\rm Tr}[\exp(-iV\tau/\hbar)\rho\otimes
|0\rangle\langle 0| \exp(iV\tau/\hbar)]$,
where ${\rm Tr}$ denotes the trace over the qubit states and $\tau$ is the 
interaction time between the TLS and the flux qubit.

When $H_{\rm env}$ is considered, the dynamics of the density operator is 
described by~\cite{optics1} 
\begin{equation}
\frac{d\rho}{dt}=r_a\ln[M(\tau)]\rho+L\rho,
\label{EOM}
\end{equation}
where $r_a$ is the rate for ``switching-on" the AA to resonantly interact with the TLS 
(each cycle includes the time required to cool the qubit) 
and $L$ describes the dissipation of the TLS due to $H_{\rm env}$. 
We model the environment in $H_{\rm env}$ by a thermal bath.
The operator $L$ can be written as~\cite{optics2}
$L\rho=-\frac{1}{2}\kappa (n_{\rm th}+1)(\sigma'_{+}\sigma'_{-}\rho
-\sigma'_{-}\rho\sigma'_{+})
-\frac{1}{2}\kappa n_{\rm th}(\sigma'_{-}\sigma'_{+}\rho
-\sigma'_{+}\rho\sigma'_{-})+{\rm H.c.}$,
where $\kappa$ is the decay rate of the TLS and $n_{\rm th}$ is the average number 
of bosons in the thermal bath (particularly, $n_{\rm th}=0$ at zero temperature).
Here we assume $g>\kappa$, ensuring coherence between the qubit and its 
ancillary circuitry.

\begin{figure}
\includegraphics[width=2.5in, 
bbllx=69,bblly=126,bburx=512,bbury=652]{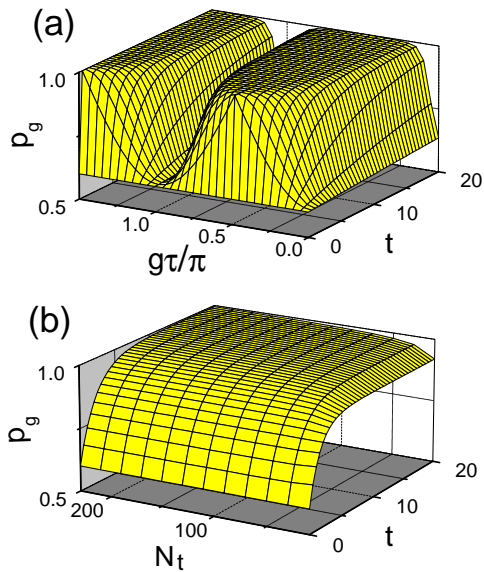}
\caption{(Color online) (a)~Ground-state probability $p_g$ versus
time $t$ (in units of $1/r_a$) and $g\tau$
for $N_t\equiv r_a/\kappa=150$. (b)~Probability $p_g$ versus $t$ 
and $N_t$ for $g\tau=0.2\pi$. Here $n_{\rm th}=0.5$ and $p_e$ is chosen to be 0.4 
at the initial time $t=0$; $g$ ($\tau$) is the interaction strength (time). 
}
\label{fig2}
\end{figure}

For the neighboring quantum TLS, Eq.~(\ref{EOM}) can be exactly solved. 
The solution for $p_e\equiv \langle e|\rho|e\rangle$ is 
\begin{equation}
p_e(t)=\left[p_e(0)-\frac{n_{\rm th}}{\Lambda}\right]\exp(-\Lambda \kappa t)
+\frac{n_{\rm th}}{\Lambda},
\end{equation} 
and $p_g\equiv \langle g|\rho|g\rangle=1-p_e$, where
$\Lambda=(2n_{\rm th}+1)-N_t\ln[\cos^2(g\tau)]$, with $N_t=r_a/\kappa$ denoting
the number of cycles for switching on the AA during the lifetime ($\equiv 1/\kappa$) 
of the TLS. Because of the coupling to the qubit, the decay rate is now scaled by 
a factor $\Lambda$. Clearly, $p_e=n_{\rm th}/\Lambda$ 
and $p_g=1-n_{\rm th}/\Lambda$ at steady state. 

Because $\Lambda$ is a periodic function of $g\tau$, both $p_e$ and $p_g$ 
are also periodic; e.g., at $g\tau=(2n-1)\pi/2$, with $n=1,2,\dots$, 
$\Lambda\longrightarrow +\infty$ and $p_g$ abruptly changes to $p_g=1$;
at $g\tau=n\pi$, with $n=0,1,\dots$, $\Lambda=2n_{\rm th}+1$ and $p_g$ slowly 
approaches $p_g=1-n_{\rm th}/(2n_{\rm th}+1)$. 
These features are clearly shown in Fig.~\ref{fig2}(a) for $p_g$ with $N_t=150$.   
To implement an efficient cooling, a smaller $\tau$ is desirable,
so we can only focus on the region $g\tau\in [0,\pi/2]$. 
Figure~\ref{fig2}(b) shows the time evolution of $p_g$ as a function of $N_t$ 
for $g\tau=0.2\pi$. Though $g\tau$ is away from $g\tau=\pi/2$, one can 
still drastically cool the TLS by evolving $p_g(t)$ to $p_g\sim 1$ with a large $N_t$.

{\it Cooling a quantum resonator.}{\bf---} When the system connected to the flux qubit
is a quantum resonator, the total Hamiltonian becomes
$H_t=H_q+H_{\rm res}+V+H_{\rm env}$, where $H_{\rm res}=\hbar\omega a^{\dag}a$ 
describes the quantum resonator, $V=-\hbar g(\sigma_{+}a+{\rm H.c.})$ is the 
interaction between them, and 
$H_{\rm env}$ describes all the degrees of freedom in the environment and 
their coupling to the quantum resonator. Also, we assume that when cooling the 
quantum resonator, the flux qubit is tuned in resonance to it. 

For the quantum resonator coupled to the flux qubit as well as to a thermal bath, 
the dynamics of the density operator of the quantum resonator is 
also described by Eq.~(\ref{EOM}). 
The operator $L$ describes the dissipation of the quantum resonator induced by
the thermal bath~\cite{optics2}:
$L\rho=-\frac{1}{2}\kappa (n_{\rm th}+1)(a^{\dag}a\rho+ \rho a^{\dag}a
-2a\rho a^{\dag})
-\frac{1}{2}\kappa n_{\rm th}(aa^{\dag}\rho + \rho aa^{\dag}-2a^{\dag}\rho a)$,
where $\kappa$ is the damping rate of the quantum resonator and 
$n_{\rm th}$ is the average number of bosons in the thermal bath coupled to 
the quantum resonator. In the present case, Eq.~(\ref{EOM}) 
can only be solved approximately.
Here we use $\ln[M(\tau)]\approx (M-1)-\frac{1}{2}(M-1)^2$, which corresponds to 
neglecting terms of order $O(\sin^6(g\tau\sqrt{n}))$. 
The equation of motion for the boson number distribution 
$p_n=\langle n|\rho|n\rangle$ of the quantum resonator becomes 
\begin{eqnarray}
\frac{dp_n}{dt}\!&\!=\!&\!a_{n+1}p_{n+1} - b_{n+1}p_{n+2} - c_{n+1}p_{n} \nonumber\\
&&\!-a_np_n + b_np_{n+1} + c_np_{n-1}, 
\end{eqnarray}
with $a_n=r_aS(n)[1+\frac{1}{2}S(n)]+\kappa (n_{\rm th}+1)n$,
$b_n=\frac{1}{2}S(n)S(n+1)$, and $c_n=\kappa n_{\rm th}n$,
where $S(n)=\sin^2(g\tau\sqrt{n})$.

At steady state, $dp_n/dt=0$, which leads to a recursion relation for the steady
boson number distribution $p_n$:
\begin{eqnarray}
&&p_{n-1}=p_n \left\{\frac{n_{\rm th}+1}{n_{\rm th}}
+\frac{N_t S(n)[2+S(n)]}
{2nn_{\rm th}}\right\}  \nonumber\\
&&~~~~~~~~~~ -p_{n+1}\frac{N_t S(n)S(n+1)}
{2nn_{\rm th}}  \,,
\label{RECUR}
\end{eqnarray}
where $N_t=r_a/\kappa$ represents the number of cycles for switching on the AA 
during the lifetime of the quantum resonator. For $N\gg 1$, 
$p_N(n_{\rm th}+1)\gg  p_{N+1}N_t\,S(N)S(N+1)/2N$, 
so we approximately have 
$p_{N-1}=\left\{\frac{n_{\rm th}+1}{n_{\rm th}}
+\frac{N_t S(N)[2+S(N)]}{2Nn_{\rm th}}\right\}p_N$.
This is the initial condition for Eq.~(\ref{RECUR}) and $p_N$ is determined by    
$\sum_{n=0}^{N}p_n=1$.

\begin{figure}
\includegraphics[width=2.5in, 
bbllx=68,bblly=154,bburx=505,bbury=671]{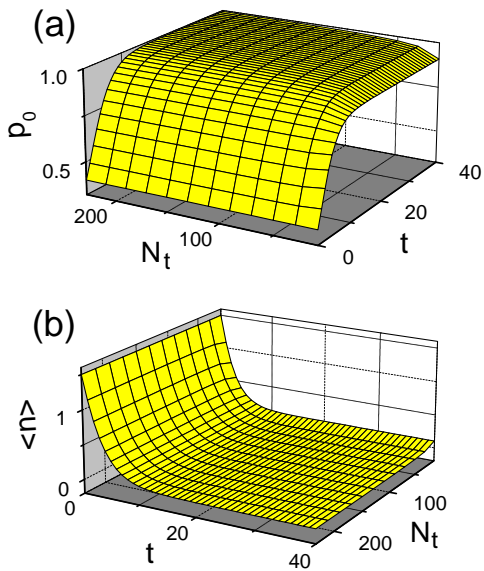}
\caption{(Color online) (a)~Vacuum-state probability $p_0$ versus
time $t$ (in units of $1/r_a$) and $N_t\equiv r_a/\kappa$. 
(b)~Average boson number $\langle n\rangle$ versus $t$ 
and $N_t$. Here $n_{\rm th}=0.5$, $g\tau=0.2\pi$, and 
$\langle n\rangle$ is chosen to be $1.5$ at the initial time $t=0$. 
}
\label{fig3}
\end{figure}

Figure~\ref{fig3} displays the time evolution of both vacuum-state probability $p_0$ 
and average boson number $\langle n\rangle$ as a function of $N_t$ for the quantum 
resonator, where $g\tau=0.2\pi$. As shown in Fig.~\ref{fig3}(a), 
one can evolve $p_0(t)$ 
to $p_0\sim 1$ using a large $N_t$. Figure~\ref{fig3}(b) shows that 
$\langle n\rangle\sim 0$ when $p_0\sim 1$, revealing that the quantum 
resonator can also be effectively cooled. 
More interestingly, Figs.~\ref{fig2}(b) and \ref{fig3}(a) give quite similar results, 
although very different models are used for the quantum systems connected to the qubit. 
This reveals that the cooling is robust and effective, irrespective of the chosen 
neighboring quantum systems.

{\it Discussion and conclusion.}{\bf---}The cooling approach studied here has 
potentially wide applications. For instance,
the environmental noise is sometimes explained as mainly due to two-level 
fluctuators, in which one or a few fluctuators play a dominant role. 
Also, the environment is often modeled by a boson bath, in which the 
bosons in resonance to the qubit play a dominant role. 
Here the quantum TLS can be used to model a two-level fluctuator 
and the quantum resonator can be used to model the dominant bosons of the environment 
in resonance to the qubit. 
Actually, the TLS defect that is most strongly coupled to the qubit may be 
off resonant to the qubit. If the off-resonance is large, the effect of the TLS 
on the qubit is not important. Otherwise, to cool the TLS defect, one can vary 
the reduced flux $f$ to tune the qubit to be in resonance with the defect. 
Also, when the environmental bosons in resonance with the qubit are cooled, 
one can tune the qubit by changing $f$ to further cool the off-resonant bosons.
After cooling the dominant noise sources of the qubit, 
the quantum coherence of the cooled flux qubit will be enhanced. 
The quantum TLS can also model a solid-state qubit and the approach can be used to 
describe cooling two coupled qubits. Naturally, the quantum resonator can 
model a mechanical resonator at the nanometer scale. The cooling of mechanical 
resonators is currently a popular topic and its study provides opportunities 
to observe the transition bewteen classical and quantum behaviors of a 
mechanical resonator~\cite{mechanical}. 
In our proposal, the quantum states can be manipulated quickly, due to the 
advantages of the proposed solid-state three-level system. 
Moreover, the cooling of both the flux qubit and the mechanical resonator can 
simultaneously enhance the quantum behaviors of the two subsystems. 
This will help observe the transition between classical and quantum behaviors of the 
mechanical resonator via measuring the quantum states of the qubit.

In conclusion, we have proposed an approach to simultaneously cool
a flux qubit and its neighboring quantum system. 
In each cycle of cooling, the flux qubit is first prepared to the ground state,
following an inverse process of the state population inversion, and 
then switched on to resonantly interact with the neighboring quantum system.  
As typical examples, we model the quantum system connected to the qubit by either 
a TLS or a resonator. Our results show that the cooling is robust and effective, 
irrespective of the chosen quantum systems. 

We thank S. Ashhab and H.-S. Goan for useful discussions. This work
was supported in part by the NSA, LPS, ARO, and the NSF grant No.~EIA-0130383.
J.Q.Y. was supported by the NSFC
grant Nos.~10474013 and 10534060,
and the NFRPC grant No.~2006CB921205.


\end{document}